\documentclass[12pt, onecolumn]{IEEEtran}

\usepackage{cite}
\usepackage{graphicx}
\usepackage{epsfig,psfrag}
\usepackage{array}
\usepackage{theorem}
\usepackage{amsmath}
\usepackage{amssymb}
\usepackage{tikz}
\usepackage{pstricks}

\newtheorem{theorem}{Theorem}

\newtheorem{lemma}{Lemma}

\newtheorem{definition}{Definition}

\newtheorem{assumption}{Assumption}

\newcommand{\mc}{\mathcal}
\newcommand{\mb}{\mathbf}

\newcommand{\LL}{\text{Ucomp}}
\newcommand{\LLE}{\text{UcompE}}
\newcommand{\LLD}{\text{UcompD}}
\newcommand{\LLED}{\text{UcompED}}

\newcommand{\MM}{m}

\begin{document}
%
\title{Universal Compression with Side Information from a Correlated  Source}
\author{Ahmad~Beirami and Faramarz Fekri

\thanks{
The material in this paper was presented in part at the 2012 IEEE International Symposium on Information Theory (ISIT 2012)~\cite{ISIT12_gain} and the 2014 IEEE Information Theory Workshop (ITW 2014)~\cite{ITW14}.}

\thanks{This work was supported by the National Science Foundation under Grant No. CNS-1017234.}

\thanks{A. Beirami was with the School
of Electrical and Computer Engineering, Georgia Institute of Technology, Atlanta,
GA, 30332-0250 USA. He is currently with EA Digital Platform -- Data \& AI, Electronic Arts, Redwood City, CA. e-mail: beirami@ece.gatech.edu.}

\thanks{F. Fekri is with the School
of Electrical and Computer Engineering, Georgia Institute of Technology, Atlanta,
GA, 30332-0250 USA e-mail: fekri@ece.gatech.edu.}
}

\maketitle

\begin{abstract}
 Packets originated from an information source in the network can be highly correlated. These packets are often routed through different paths, and compressing them requires to process them individually.
Traditional universal compression solutions would not perform well over a single packet because of the limited data available for learning the unknown source parameters.
In this paper, we define a notion of correlation between information sources and characterize the average redundancy in universal compression with side information from a correlated source.
 We define the side information gain as the ratio between the average maximin redundancy of universal compression without side information to that with side information.
 We derive a lower bound on the side information gain, where we show that the presence of side information provides at least $50\%$ traffic reduction over traditional universal compression when applied to network packet data confirming previous empirical studies.
\end{abstract}


\section{Introduction}
\label{sec:intro}

Several studies have inferred the presence of considerable amount of correlation in network traffic data~\cite{SIVAREP,anand_sigcomm_08, anand_sigcomm_09, anand_sigmetrics_09,Spring2000}.
Specifically, we may broadly define correlation in two dimensions: 
\begin{enumerate}
\item Temporal correlation within content from an information source being delivered to a client. 
\item Spatial correlation across content from different information sources delivered to the same/different clients.
\end{enumerate}
Network traffic abounds with the first dimension of temporal correlation, which is well understood.
For example, if traffic contains mostly English text, there is significant correlation within the content.  
The existence of the second dimension of correlation is also confirmed in several real data experiments~\cite{SIVAREP,anand_sigcomm_08, anand_sigcomm_09, anand_sigmetrics_09, Spring2000}.

This has motivated the employment of correlation elimination techniques for network traffic data.\footnote{Within the networking community, these are known as correlation elimination (RE) techniques but since redundancy has a specific meaning within the universal compression community, as shall be formally defined in the sequel, we chose to refer to these techniques as correlation elimination for the clarity of discussion.}
The present correlation elimination techniques are mostly based on content caching mechanisms used by solutions such as web-caching~\cite{web-cache}, CDNs~\cite{CDN}, and P2P networks~\cite{P2P-app}. However, caching approaches that take place at the application layer, do not effectively leverage the spatial correlation, which exists mostly at the packet level~\cite{SIVAREP,anand_sigcomm_08, anand_sigcomm_09, anand_sigmetrics_09}. To address these issues, a few studies have considered ad-hoc methods such as packet-level correlation elimination (deduplication) in which redundant transmissions of segments of a packet that are seen in previously sent packets are avoided~\cite{anand_sigcomm_09, anand_sigmetrics_09}. However, these techniques are limited in scope and can only eliminate exact duplicates from the segments of the packets leaving statistical correlations intact.

It is natural to consider universal compression algorithms for correlation elimination from network traffic data. 
While universal compression algorithms, e.g., Lempel-Ziv algorithms~\cite{LZ77,LZ78} and context tree weighting (CTW)~\cite{CTW95}, have been very successful in many domains, they do not perform very well on limited amount of data  as learning the unknown source statistics imposes an inevitable redundancy (compression overhead). This redundancy depends on the richness of the class of the sources with respect to which the code is universal~\cite{Davisson_noiseless_coding,Rissanen_1984, Merhav_Feder_IT, Barron_MDL, ISIT11}.
Further, traditional universal compression would only attempt to deal with temporal correlation from a stationary source and lacks the structure to leverage the spatial correlation dimension.

In this paper, as an abstraction of correlation elimination from network traffic, we study universal compression with side information from a correlated source. The organization of the paper and our contributions are summarized below. 
\begin{itemize}
\item In Section~\ref{sec:motivation}, we demonstrate that universal compression of finite-length sequences (up to hundreds of kilobytes) fundamentally suffers from a significant compression overhead. This motivates using side information for removing this redundancy.
\item In Section~\ref{sec:setup}, we present the formal problem setup. We define a notion of correlation between two parametric information sources, and study strictly lossless and almost lossless compression when side information from a correlated source is available to the encoder and/or the decoder.
\item In Section~\ref{sec:correlation}, we establish several nice properties of correlated information sources. We show that the degree of correlation is tuned with a single hyperparameter, which results in independent information sources in one end and duplicate sources in the other end.
\item In Section~\ref{sec:redundancy}, we characterize the average maximin redundancy with side information from a correlated  source. We also show that if permissible error is sufficiently small the redundancy of almost lossless compression dominates the reduction in codeword length due to the permissible error.
\item In Section~\ref{sec:gain}, we define and characterize a notion of side information gain and establish a sufficient condition on the length of a side information string that would guarantee almost all of the benefits. We show that the side information gain can be considerable in many scenarios and derive a cutoff threshold on the size of memory needed to obtain all of the side information gain.
\item In Section~\ref{sec:one-to-one}, we show that the side information gain is largely preserved even if the prefix constraint on the code is dropped.
\item In Section~\ref{sec:network}, we provide a case study that shows how these benefits would be extended in a network setting.
\item Finally, the conclusions are summarized in Section~\ref{sec:conclusion}.
\end{itemize}

\begin{figure}
\centering
\includegraphics[height=1.2in]{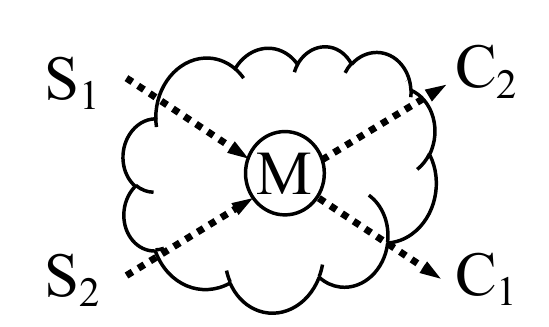}
\caption{The basic abstraction of universal compression with side information.}
\label{fig:single_basic}
\end{figure}

\section{Motivation}
\label{sec:motivation}
We describe universal compression with side information from a correlated source in the most basic scenario. We use the notation $x^n = (x_1,...,x_n)$ to denote a string of length $n$ on the finite alphabet $\mc{X}$. For example, for an 8-bit alphabet that has 256 characters, each $x_i$ is a byte and $x^n$ denotes a packet at the network layer. We assume that, as shown in Fig.~\ref{fig:single_basic}, the network consists of content server nodes $S_1$ and $S_2$, an intermediate memory-enabled (relay or router) node $M$, and client nodes $C_1$ and $C_2$. 

Let's assume that the content at $S_1$ is stationary and correlated with the content at $S_2$. Assume that $y^m$ has already been routed through $S_2 \to M\to C_2$ path. Also, assume that all nodes in the route, i.e., $S_2$, $M$ and $C_2$, have memorized the content $y^m$.
Now, assume that $x^n$ is to be routed through $S_1 \to M \to C_1$ path. In this case, at the $S_1 \to M$ link the side information string is only available to the decoder, while at the $M \to C_1$ link, the side information is only available to the encoder. If $x^n$ was to be routed through $S_1 \to M \to C_2$ path, in this case, the side information would be available to both the encoder and the decoder at $M \to C_2$ link.
As such, we wish to study universal compression with side information that is available to the encoder and/or the decoder in this paper.

Given the side information gain, in~\cite{TON14}, we analyzed the network-wide benefits of introducing memory-enabled nodes to the network and provided results on memory placement and routing for extending the gain to the entire network. However,~\cite{TON14} did not explain how to characterize the side information gain.

Let redundancy be the overhead in the number of bits used for describing a random string drawn from an unknown information source compared to the optimal codeword length given by the Shannon code.
In the universal compression of a family of information sources that could be parametrized with $d$ unknown parameters, Rissanen showed that the expected redundancy asymptotically scales as $\frac{d}{2}\log n + o(\log n)$ for almost all sources in the family~\cite{Rissanen_1984}.\footnote{$f(n) = o(g(n))$ if and only if $\lim_{n\to \infty}\frac{f(n)}{g(n)} = 0.$} Clarke and Barron~\cite{Clarke_Barron}  derived the asymptotic average minimax redundancy for memoryless sources to be $\frac{d}{2} \log n + O_n(1)$.\footnote{$f(n) =O(g(n))$ if and only if $\lim_{n\to \infty} \sup \frac{f(n)}{g(n)} < \infty$.} This was later generalized by Atteson to Markov information sources~\cite{atteson_markov}. The average minimax redundancy is concerned with the redundancy of the worst parameter vector for the best code, and hence, does not provide much information about the rest of the source parameter values. However, in light of Rissanen's result one would expect that asymptotically almost all information sources in the family behave similarly. The question remains as how these  would behave in the finite-length regime.

In~\cite[Theorem 1]{ISIT11}, using a probabilistic treatment, we derived sharp lower bounds on the probability of the event that the redundancy in the compression of a random string of length $n$ from a parametric source would be larger than a certain fraction of $\frac{d}{2}\log n$.
\cite[Theorem 1]{ISIT11} provides, for any $n$, a lower bound on the probability measure of the information sources for which the average redundancy of the best universal compression scheme would be larger than ${\frac{d}{2}\log n}$. To demonstrate the implications of this result in the finite-length regime of interest in this paper, we consider an example using a first-order Markov information source with alphabet size $k=256$.  This information source is represented using $d = 256 \times 255 = 62580$ parameters. We further assume that the source entropy rate is $0.5$ bit per byte ($H^n(\theta)/n = 0.5$). This assumption is inspired by experiments on real network data traffic in~\cite[Section IV.A]{TON14}.
It is implied by \cite[Theorem 1]{ISIT11}  that the compression overhead is more than $75\%$ for strings of length $256$kB. We conclude that redundancy is significant in the compression of finite-length low-entropy sequences, such as the Internet traffic packets that are much shorter than $256$kB. It is this redundancy that we hope to suppress using side information from a correlated source. The compression overhead becomes negligible for very long sequences (e.g., it is less than $2\%$ for strings of length $64$MB and above), and hence, the side information gain vanishes as $O\left(\log n/n\right)$ when the sequence length grows large.

It is also worth noting the scope of benefits expected from universal compression of network traffic with side information is significant since file sharing and web data comprise more than $50\%$ of network traffic~\cite{cisco-vni} for which, the correlation levels may be as high as $90\%$~\cite{SIVAREP}. Further, universal compression with side information is applicable to storage reduction in cloud and distributed storage systems, traffic reduction for Internet Service Providers, and power and bandwidth reduction in wireless communications networks (e.g., wireless sensors networks, cellular mobile networks, hot spots). See~\cite{TON14, TWC15} for a more thorough investigation of such applications and also for practical coding schemes for network packet compression.

\section{Problem Setup}
\label{sec:setup}
Let $\mc{X}$ be a finite alphabet. We assume that the server $S$ comprises of two parametric sources $\theta^{(1)}$ and $\theta^{(2)}$, which are defined using parameter vectors $\theta^{(1)} = (\theta^{(1)}_1,...,\theta^{(1)}_d)$ and $\theta^{(2)} = (\theta^{(2)}_1,...,\theta^{(2)}_d)$, where $\theta^{(1)}, \theta^{(2)} \in \Theta_d$ and $\Theta_d$ is a $d$-dimensional set. Denote $\mu^n_{\theta^{(1)}}$ and $\mu^n_{\theta^{(2)}}$ as the probability measures defined by the parameter vector $\theta$ on strings of length $n$. 
If the information sources are memoryless, $\theta^{(1)}$ would denote the categorical distribution of the characters, and $\mu^n_{\theta^{(1)}}$  would be a product distribution.
We assume that $\theta^{(1)}$ is a priori unknown.
Unless otherwise stated, we use the notation $X^n \in \mc{X}^n$  and $Y^m \in \mc{X}^m$ to denote random string of length $n$ and $m$ drawn from $\mu^n_{\theta^{(1)}}$ and $\mu^m_{\theta^{(2)}}$, respectively.
See Assumption~\ref{assump:reg} (appendix) for a set of regularity conditions that we assume on the parametric family.

\begin{figure}
\begin{center}
\includegraphics[width = 0.35\textwidth]{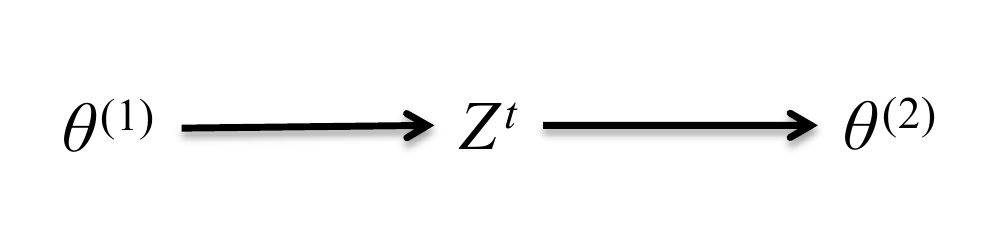}
\vspace{-.3in}
\end{center}
\caption{The correlation model between the two information sources.}
\label{fig:correlation_model}
\end{figure}

We put forth a notion of correlation between the parameter vectors $\theta^{(1)}$ and $\theta^{(2)}$, where the correlation could be tuned using a hyperparameter $t$.
We assume that  the unknown (and unobserved) parameter vector $\theta^{(1)}$ follows a prior distribution $q$ supported on $\Theta_d$. Let $Z^{t}$ be a random string of length $t$ that is drawn from $\mu^t_{\theta^{(1)}}$.
We  assume that given $Z^{t}$, the parameter vectors $\theta^{(1)}$ and $\theta^{(2)}$ are independent and identically distributed. This is shown in the Markov chain represented in Fig.~\ref{fig:correlation_model}.  We will state several nice properties of this proposed model in Section~\ref{sec:correlation}.

Note that this framework is fundamentally different from Slepian-Wolf coding that also targets the spatial correlation between distributed information sources~\cite{slepian_wolf,Yang-SW,Mina_TCOM,PRADHAN-DSC}. In  Slepian-Wolf coding, the sequences from the distributed sources are assumed to have character-by-character correlation, which is also different from our correlation model that is due to the parameter vectors being unknown in a universal compression setup.

 Let $H^n(\theta)$ denote the Shannon entropy of the source given $\theta$, i.e.,
\begin{align}
H^n(\theta) &\triangleq  E \left\{ \log \left(\frac{1}{\mu_\theta(X^n)} \right) \right\}\nonumber \\
&=\sum_{x^n \in \mc{X}^n}\mu_{\theta}(x^n) \log \left(\frac{1}{\mu_\theta(x^n)}\right).
\end{align}
Throughout this paper expectations are taken over functions of the random sequence $X^n$ with respect to the (unknown) probability measure $\mu_\theta$,  and $\log(\cdot)$  denotes the logarithm in base $2$, unless otherwise stated.
We further use the notation $H(\theta)$ to denote the entropy rate, defined as $
H(\theta) \triangleq \lim_{n \to \infty }\frac{1}{n} H^n(\theta).
$

Let $\mc{I}(\theta)$ be the Fisher information matrix, where each element is given by
\begin{equation}
\mc{I}(\theta)_{ij}\triangleq \lim_{n \to \infty} \frac{1}{n \log e}E\left\{\frac{\partial^2}{\partial \theta_i \partial \theta_j} \log\left( \frac{1}{\mu^n_\theta(X^n)}\right)\right\}.
\label{eq:fisher}
\end{equation}
Fisher information matrix quantifies the amount of information, on the average, that a random string $X^n$ from the source conveys about the source parameters.
Let Jeffreys' prior on $\Theta_d$ be defined as
\begin{equation}
w_J(\theta) \triangleq \frac{|\mc{I}(\theta)|^{\frac{1}{2}}}{\int_{\phi \in \Theta_d} | \mc{I}(\phi)|^\frac{1}{2}d \phi}.
\label{eq:Jeffreys}
\end{equation}
Roughly speaking, Jeffreys' prior is optimal in the sense that the average minimax redundancy is asymptotically achieved when the parameter vector $\theta$ is assumed to follow Jeffreys' prior (see~\cite{Clarke_Barron} for a formalized statement and proof). This prior distribution is particularly interesting because it corresponds to the worst-case compression performance for the best compression scheme.

We consider the family of block codes that map any $n$-string to a variable-length binary sequence, which also satisfy Kraft's inequality~\cite{huffman_redundancy}.
Let 
\begin{align}
\mc{C} &\triangleq \mc{X}^n \times \{\varepsilon\}^m,\nonumber\\
\mc{C}^E &\triangleq \mc{X}^n \times \mc{X}^m,\nonumber\\
\mc{D} & \triangleq \{0, 1\}^* \times   \{\varepsilon \}^m,\nonumber\\
\mc{D}^D &\triangleq  \{0, 1\}^* \times  \mc{X} ^m,
\label{eq:mcDE}
\end{align}
where $\varepsilon$ denotes an erasure.
We use $c: \mc{C} \to \{0, 1\}^*$ and $c^E: \mc{C}^E \to \{0, 1\}^*$ to denote the encoder without and with side information, respectively. Similarly, we also use $d: \mc{D} \to \mc{X}^n$ and $d^D: \mc{D}^D \to \mc{X}^n$ to denote the decoder without and with side information, respectively. 
We use notations 
\begin{align}
C &\triangleq (\mc{X}^n , \varepsilon^m)  \in \mc{C},\nonumber\\
C^E &\triangleq (\mc{X}^n , y^m)  \in \mc{C}^E,\nonumber\\
D &\triangleq (c(C) , \varepsilon^m) \in \mc{D} ,\nonumber\\
D^E &\triangleq (c^E(C^E) , \varepsilon^m)  \in \mc{D},\nonumber\\
D^{D} &\triangleq (c(C) , y^m)  \in \mc{D}^D,\nonumber\\
D^{ED} &\triangleq (c^{E}(C^E) , y^m) \in \mc{D}^D.
\label{eq:DE}
\end{align}

Next, we present the notions of strictly lossless and almost lossless source codes, which will be needed in the sequel. While the definitions are only given for the case with no side information at the encoder and the decoder, it is straightforward to extend them using the above definitions.
Our main focus in on prefix free codes that ensure unique decodability of concatenated code blocks (see~\cite[Chapter 5.1]{cover-book}).

\begin{definition}
The code $c: \mc{C}\to \{0,1\}^*$ is called strictly lossless (also called zero-error) if there exists a reverse mapping $d: \mc{D} \to \mc{X}^n$ such that
$$\forall C \in\mc{C}:\quad d(D) = x^n.$$
\end{definition}
\begin{definition}
Let $\mu^{n,m}$ denote a joint probability distribution on $(x^n, y^m)$.
The code $c_{\epsilon}: \mc{C} \to \{0,1\}^*$ is called almost lossless with permissible error probability $\epsilon(n)$, if there exists a reverse mapping $d_{\epsilon}:\mc{D} \to \mc{X}^n$ such that
$$E\{ \mb{1}_e(X^n, Y^m) \}\leq \epsilon(n),$$
where $\mb{1}_e(x^n)$ denotes the error indicator function, i.e,
\begin{equation}
\mb{1}_e(x^n, y^m) \triangleq
\left\{
\begin{array}{ll}
1 & \quad \quad d_{\epsilon}(D) \neq x^n,\\
0 & \quad \quad \text{otherwise},
\end{array}
\right.\nonumber
\end{equation}
where $D$ and $E$ are defined in~\eqref{eq:DE}.
\end{definition}
Most of the practical data compression schemes are examples of strictly lossless codes, namely, the arithmetic code~\cite{Arithmetic_coding_introduction}, Huffman code~\cite{Huffman_coding}, Lempel-Ziv codes~\cite{LZ77,LZ78}, and CTW code~\cite{CTW95}.
In almost lossless source coding, which is a weaker notion of the lossless case, we allow a non-zero error probability $\epsilon(n)$ for any finite $n$ while if $\epsilon(n) = o_n(1)$ the code is \emph{almost surely} asymptotically error free. The proofs of Shannon~\cite{Shannon_paper} for the existence of entropy achieving source codes are based on almost lossless random codes. The proof of the Slepian-Wolf theorem~\cite{slepian_wolf} also uses almost lossless codes. Further, all of the practical implementations of SW source coding are based on almost lossless codes (see~\cite{PRADHAN-DSC,Mina_TCOM}).

\begin{figure}
\begin{center}
\includegraphics[width = 0.47\textwidth]{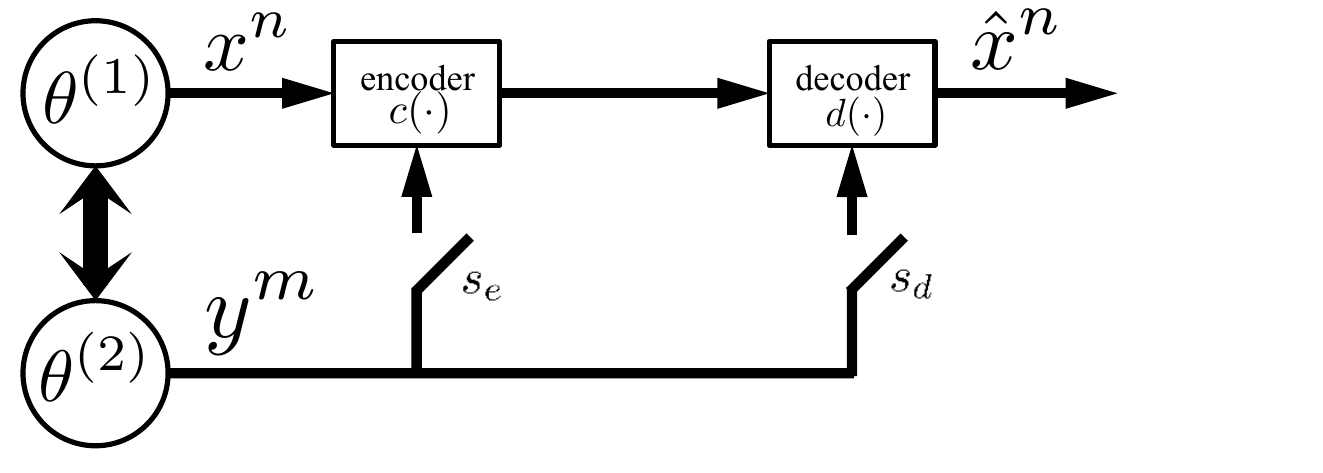}
\end{center}
\caption{The compression model for universal source coding with two correlated parameter vectors.}
\label{fig:two_model}
\end{figure}

We consider four coding strategies according to the orientation of the switches $s_e$ and $s_d$ in Fig.~\ref{fig:two_model}
for the compression of $x^n$ drawn from $\mu^n_{\theta^{(1)}}$ provided that the sequence $y^m$ drawn from $\mu^m_{\theta^{(2)}}$ is available to the encoder/decoder or not.\footnote{In this paper, we assume that $m$ and $n$ are a priori known to both the encoder and the decoder.}

\begin{itemize}
\item {{\LL}} (Universal compression without side information), where the switches $s_e$ and $s_d$  in Fig.~\ref{fig:two_model} are both {\em open}. This corresponds to $C \in \mc{C}$ and $D \in \mc{D}$.
\item {\LLE} (Universal compression with encoder side information), where the switch $s_e$ in Fig.~\ref{fig:two_model} is {\em closed} but the switch $s_d$ is {\em open}. This corresponds to $C^E \in \mc{C}^E$ and $D^E \in \mc{D}$.

\item {{\LLD}} (Universal compression with decoder side information), where the switch $s_e$ in Fig.~\ref{fig:two_model} is {\em open} but the switch $s_d$ is {\em closed}. This corresponds to $C \in \mc{C}$ and $D^D \in \mc{D}^D$.

\item { {\LLED}} (Universal compression with encoder-decoder side information), where the switches $s_e$ and $s_d$  in Fig.~\ref{fig:two_model} are both {\em closed}. This corresponds to $C^E \in \mc{C}^E$ and $D^{ED} \in \mc{D}^D$.

\end{itemize}

\section{Implications of the Correlation Model}
\label{sec:correlation}

In this section, we study some implications of the proposed correlation model. This section may be skipped by the reader and only referred to when a particular lemma is needed in the subsequent proofs.

\begin{lemma}
The joint distribution of $(\theta^{(1)}, \theta^{(2)})$  for all $t \geq 0$ is given by 
\begin{equation}
p^{t}(\theta^{(1)}, \theta^{(2)}) = w(\theta^{(1)}) w(\theta^{(2)}) f^t(\theta^{(1)}, \theta^{(2)}),
\end{equation}
where $f^t(\theta^{(1)}, \theta^{(2)})$ is defined as
\begin{equation}
f^t(\theta^{(1)}, \theta^{(2)}) \triangleq \sum_{z^{t} \in \mc{X}^{t}} \left( \frac{\mu^t_{\theta^{(1)}}(z^{t}) \mu^t_{\theta^{(2)}}(z^{t})}{ \int_{\phi \in \Theta_d} \mu^t_{\phi}(z^{t}) w(\phi) d\phi } \right).
\end{equation}
\label{lem:cond}
\end{lemma}
\begin{IEEEproof}
We have
\begin{align}
p^{t}(\theta^{(2)}|\theta^{(1)}) &= \sum_{z^{t} \in \mc{X}^{t}}p (\theta^{(2)},z^{t}|\theta^{(1)})\\
&=  \sum_{z^{t} \in \mc{X}^{t}} p(\theta^{(2)}|z^{t}) \mu^t_{\theta^{(1)}}(z^{t})\label{eq:indep}\\
&= w(\theta^{(2)}) \sum_{z^{t} \in \mc{X}^{t}} \left( \frac{\mu^t_{\theta^{(1)}}(z^{t}) \mu^t_{\theta^{(2)}}(z^{t})}{ \int_{\phi \in {\Theta_d}} \mu^t_{\phi}(z^{t}) w(\phi) d \phi } \right),\label{eq:bayes}
\end{align}
where~\eqref{eq:indep} follows from the fact that $\theta^{(2)}$ and $\theta^{(1)}$ are independent and identically distributed given $Z^{t}$, and~\eqref{eq:bayes} follows from the Bayes rule. Hence, the result follows.
\end{IEEEproof}

Next, we find the marginal distribution of $\theta^{(2)}$, i.e., $p^{t}_{\theta^{(2)}}(\theta^{(2)})$.
\begin{lemma}
For all $t \geq 0$, we have
\begin{equation}
p^{t}(\theta^{(2)}) = w(\theta^{(2)}).
\end{equation}
\label{lem:marginal}
\end{lemma}
\begin{IEEEproof}
The proof follows from the following equations
\begin{align}
p^{t}_{\theta^{(2)}}(\theta^{(2)}) &= \int_{{\Theta_d}}p^{t}_{\theta^{(2)}|\theta^{(1)}}(\theta^{(2)}|\theta^{(1)}) w(\theta^{(1)}) d\theta^{(1)}\nonumber\\
&=  w(\theta^{(2)}) \int_{\Theta_d} f^t( \theta^{(1)}, \theta^{(2)}) w(\theta^{(1)}) d\theta^{(1)}\label{eq:dd} \\
& = w(\theta^{(2)}),
\end{align}
where~\eqref{eq:dd} follows from Lemma~\ref{lem:integral} (appendix).
\end{IEEEproof}

\begin{lemma}
$\theta^{(2)}$ is independent of $\theta^{(1)}$ if and only if $t=0$, i.e., 
\begin{equation}
p^{0} (\theta^{(1)}, \theta^{(2)}) = w(\theta^{(1)}) w(\theta^{(2)}).
\end{equation}
\label{lem:independent}
\end{lemma}
\begin{IEEEproof}
By definition of $f^t(\cdot, \cdot)$, and the fact that $\mu^0_{\theta^{(1)}}(z^0) = 1$, we have
\begin{equation}
f^0( \theta^{(1)}, \theta^{(2)})=  \left( \frac{1}{ \int_{\phi \in {\Theta_d}} w(\phi) d\phi} \right) = 1.
\label{eq:f-prop2}
\end{equation}
Hence, the claim follows by invoking Lemma~\ref{lem:cond}.
\end{IEEEproof}

\begin{lemma}
$\theta^{(2)}$ converges in mean square to $\theta^{(1)}$ as $t \to \infty$, that is 
\begin{equation}
\lim_{t \to \infty} E \left\{ \lVert\theta^{(2)} - \theta^{(1)} \rVert^2 \right\}= 0.
\end{equation}
\label{lem:conv-mean-sq}
\end{lemma}
\begin{IEEEproof}
Let $\widehat{\theta^{(1)}}(Z^{t})$ be the maximum likelihood estimator (MLE) of $\theta^{(1)}$ from the observation $Z^{t}$. By definition, $\widehat{\theta^{(1)}}(Z^{t})$ also serves as the MLE for $\theta^{(2)}$.
Then,
\begin{align}
E \{\lVert \theta^{(2)} - \theta^{(1)} \rVert^2\}  & \leq E \{\lVert\theta^{(2)} -\widehat{\theta^{(1)}}(Z^{t}) \rVert^2 \} \nonumber\\
&+ E\{ \lVert\theta^{(1)}- \widehat{\theta^{(1)}}(Z^{t}) \rVert^2 \}\\
& = 2 E\{ \lVert\theta^{(1)}- \widehat{\theta^{(1)}}(Z^{t}) \rVert^2\},
\end{align}
and the statement follows from the convergence of MLE in mean square for the parametric information source as assumed in the regularity conditions put forth in Assumption~\ref{assump:reg} (appendix).
\end{IEEEproof}

{\em Remark:} The degree of correlation between the two parameter vectors $\theta^{(1)}$  and $\theta^{(2)}$ is determined by the hyperparameter $t$.
This degree of correlation varies from independence of the two parameter vectors at $t=0$ all the way to the vectors being equal (convergence in mean square) when $t \to \infty$. Further note that the covariance matrix of the parameter vectors $\theta^{(1)}$ and $\theta^{(2)}$ asymptotically as $t$ grows large behaves like $ \frac{2}{{t}} \mc{I}^{-1} (\theta^{(1)})$.

\section{Average Maximin Redundancy}
\label{sec:redundancy}
In this section, we investigate the average maximin redundancy in universal compression of correlated sources for different coding strategies put forth in Section~\ref{sec:setup}.

\subsection{{\LL} Coding Strategy}
Let $l^n :\mc{X}^n \to \mathbb{R}^+ $ denote the universal (strictly lossless) length function for {\LL}.\footnote{We ignore the integer constraint on the length functions in this paper, which will result in a negligible redundancy smaller than $1$ bit and is exactly analyzed in~\cite{Precise_Minimax_Redundancy, huffman_redundancy}.} 
This is the length associated with a strictly lossless code. A necessary and sufficient condition for existence of a code that satisfies unique decodability is given by Kraft inequality:
\begin{equation}
\sum_{x^n \in \mc{X}^n} 2^{-l^n(x^n)} \leq 1.
\label{eq:Kraft}
\end{equation}
Denote $L^n$  as the set of all strictly lossless universal length functions that satisfy Kraft inequality.
Denote $R^n(l^n,\theta)$ as the average redundancy of the code with length function $l^n(\cdot)$, defined as
\begin{equation}
R^n(l^n,\theta) \triangleq E_{X^n \sim \mu^{n}} \{ l^n(X^n) \}-  H^n(\theta).
\label{eq:redundancy-def}
\end{equation}
Define $\underline{R}$ as the minimax redundancy of {\LL}, i.e., 
\begin{equation}
\underline{R}^n = \max_{w \in \Omega_d} \min_{l^n \in L^n} \int_{\theta \in \Theta_d} R^n(l^n,\theta) w(\theta)d\theta.
\label{eq:avg_min_redundancy}
\end{equation}
It is well known that the maximum above is attained by Jeffreys' prior in the asymptotic limit as $n$ grows large. Hence, in the rest of this paper we assume that $\theta^{(1)}, \theta^{(2)} \sim w_J$ follow Jeffreys' prior given in~\eqref{eq:Jeffreys}. On the other hand, the length function that achieves the inner minimization is simply the information random variable.
\begin{equation}
\imath^n(x^n) \triangleq - \log\left(\int_{\theta \in \Theta_d} \mu_\theta^n(x^n) w_J(\theta)d\theta\right).
\end{equation}
Putting it all together, we have 
\begin{align}
\underline{R}^n = I(X^n; \theta),
\end{align}
where $I(\cdot; \cdot)$ denotes the mutual information. This is Gallager's redundancy-capacity theorem in~\cite{Gallager-source-coding}.

Clarke and Barron~\cite{Clarke_Barron} showed that the average maximin redundancy for strictly lossless {\LL} is
\begin{equation}
\underline{R}^n = \frac{d}{2}  \log\left( \frac{n}{2\pi e} \right) + \log \int_{\phi \in \Theta_d}
|\mc{I}(\phi)|^{\frac{1}{2}}d\phi + o_n(1).\nonumber
\label{eq:maximin}
\end{equation}
This result states that the average maximin redundancy in {\LL} coding strategy is $O(\log n)$ and also is linearly proportional to the number of unknown source parameters, $d$.

It is straightforward to define $\underline{R}_\epsilon^n$ as the average redundancy when $\theta^{(1)}$ follows Jeffreys' prior when we are restricted to almost lossless codes with permissible error $\epsilon$. Note that it is clear that $\underline{R}_\epsilon^n \leq \underline{R}^n $. A natural question that arises is how much reduction is achievable by allowing a permissible error probability in decoding.
Our main result on {\LL} coding strategy with almost lossless codes is given in the following theorem.
\begin{theorem}
If $\epsilon(n) = O(\frac{1}{n})$, then 
\begin{equation}
\underline{R}_\epsilon^n  = \underline{R}^n - O_n(1).
\end{equation}
\label{thm:almost-lossless}
\end{theorem}
\begin{IEEEproof}
The proof is completed by invoking Lemma~\ref{thm:NM_almost} in the appendix and noting that $\underline{R}^n = O(\log n)$.
\end{IEEEproof}

The content of Theorem~\ref{thm:almost-lossless} is that if the permissible error, $\epsilon(n)$, in almost lossless compression vanishes fast enough as $n$ grows, then asymptotically the maximin risk imposed by universality of compression dominates any savings obtained by allowing an $\epsilon(n)$ average error in decoding. Hence, in the rest of this paper we only focus on the family of strictly lossless codes.

\subsection{{\LLE} Coding Strategy}

Since the side information sequence $y^m$ is not available to the decoder, then the minimum number of average bits required at the decoder to describe the random sequence $X^n$ is indeed $H(X^n)$. On the other hand, it is straightforward to see that 
\begin{equation}
H(X^n) = H^n(\theta^{(1)}) + I(X^n;\theta^{(1)}),
\end{equation}
where
$
I(X^n;\theta^{(1)}) = \underline{R}^n
$
by the redundancy-capacity theorem.
Hence, in {\LLE} strategy, we establish that the side information provided by $y^m$ only at the encoder does not provide any benefit on the strictly lossless universal compression of the sequence $x^n$.

\subsection{{\LLD} Coding Strategy}

Considering the {\LLD} strategy, by Assumption~\ref{assump:reg} (appendix), the two sources $\mu_{\theta^{(1)}}$ and $\mu_{\theta^{(2)}}$ are $d$-dimensional parametric ergodic sources. In other words, any pair $(x^n, y^m) \in \mc{X}^n \times \mc{X}^m$ occurs with non-zero probability and the support set of $(x^n,y^m)$ is equal to the entire $\mc{X}^n \times \mc{X}^m$. Therefore, the knowledge of the side information sequence $y^m$ at the decoder does not rule out any possibilities for $x^n$ at the decoder. Hence, we conclude that side information provides no reduction in average codeword length (see~\cite{Alon_source_coding} and the references therein for a discussion on zero-error coding). However, this is not the case in almost lossless source coding. See~\cite{TWC15} for an almost lossless code in this case.

\subsection{{\LLED} Coding Strategy}

In the case of {\LLED}, let $l^{n,m} : \mc{X}^n \times \mc{X}^m \to \mathbb{R}^+$ denote the universal prefix-free strictly lossless length function. Denote $L^{n,m}$ as the set of all possible such length functions. Denote $R^{n,m}(l^{n,m}, \theta^{(1)}, \theta^{(2)})$ as the expected redundancy of the code with length function $l^{n,m}(\cdot, \cdot)$:
\begin{align}
&R^{n,m}(l^{n,m},\theta^{(1)}, \theta^{(2)}) \nonumber\\
& \triangleq  E_{X^n, Y^m \sim \mu^{n,m}_{\theta^{(1)}, \theta^{(2)}}} \{ l^{n,m}(X^n, Y^m) \}-  H^n(\theta^{(1)}),
\end{align}
where $\mu^{n,m}_{\theta^{(1)}, \theta^{(2)}}$ is the product distribution
\begin{equation}
\mu^{n,m}_{\theta^{(1)}, \theta^{(2)}} (x^n, y^m) \triangleq \mu^n_{\theta^{(1)}}(x^n) \mu^m_{\theta^{(2)}}(y^m) .
\end{equation}
Here we assume that $\theta^{(1)}, \theta^{(2)}$ follow the correlation model that we put forth in this paper with their marginals being Jeffreys' prior. Hence, we define
\begin{equation}
\underline{R}^{n,m,t} \triangleq E_{(\theta^{(1)}, \theta^{(2)}) \sim p^t } \left\{ \min_{l^{n,m} \in L^{n,m}} R^{n,m}(l^{n,m},\theta^{(1)}, \theta^{(2)}) \right\}.
\end{equation}
In this case, following the same lines of arguments in~\cite{Gallager-source-coding}, we can conclude that 
\begin{equation}
\underline{R}^{n,m,t} = I(X^n; \theta^{(1)} | Y^m).
\end{equation}

The following intuitive inequality demonstrates that the redundancy decreases when side information is available.
\begin{lemma} For all $n,m, t \geq 0$, we have
\begin{equation}
\underline{R}^{n,m,t} \leq \underline{R}^n.
\end{equation}
with equality if and only if $\min\{n, m, t\} = 0.$
\label{lem:sandwich}
\end{lemma}
\begin{IEEEproof}
First notice that $\underline{R}^{n,m,t} = I(X^n; \theta^{(1)} |Y^m)$ and $ \underline{R}^n = I(X^n; \theta^{(1)})$ and hence the inequality is achieved by applying Lemma~\ref{lem:Markov-MI} (appendix) and noticing the Markov chain $X^n \to \theta^{(1)} \to Y^m$. 

Equality holds if and only if $I(X^n; Y^m) = 0$.
We just need to show that $I(X^n; Y^m) = 0$ if and only if $\min\{n, m, t\} = 0.$
If $n = 0$ or $m = 0$, then $I(X^n; Y^m) = 0$.
If $t = 0$, then $\theta^{(1)}$ and $\theta^{(2)}$  are {\em independent} by Lemma~\ref{lem:independent}. Hence, $X^n$ and $Y^m$ are also independent. Conversely, assume that $n,m > 0$, then by Lemma~\ref{lem:independent}, $I(X^n; Y^m) = 0$ only if $t =0$ completing the proof.
\end{IEEEproof}

According to Lemma~\ref{lem:sandwich}, side information cannot hurt, which is intuitively expected. However, there is no benefit provided by the side information when the two parameter vectors of the sources $S_1$ and $S_2$ are independent. This is not surprising as when $\theta^{(1)}$ and $\theta^{(2)}$ are independent, then $X^n$ (produced by $S_1$) and $Y^m$ (produced by $S_2$) are also independent.
Thus, the knowledge of $y^m$ does not affect the distribution of $x^n$. Hence, $y^m$ cannot be used toward the reduction of the codeword length for $x^n$.

Next, we present our main result on the average maximin redundancy for strictly lossless {\LLED} coding.
\begin{theorem}
For strictly lossless {\LLED} coding, if $\min\{m, t\} = O_n(1)$, then
$$\underline{R}^{n,m,t} = \underline{R}^n - O_n(1),$$
and if $\min\{m, t\} = \omega_n(1)$, then\footnote{$f(n) = \omega(g(n))$ if and only if $g(n) = o(f(n))$.} 
\begin{equation}
\underline{R}^{n,m,t} =  \widehat{R}(n,m,t) + o_n(1),\nonumber
\end{equation}
where $\widehat{R}(n,m,t)$  is defined as
\begin{equation}
\widehat{R}(n,m,t) \triangleq  \frac{d}{2} \log\left( 1+\frac{n}{m^\star(m, t)}\right),
\label{eq:Rhat}
\end{equation}
and $m^\star(\cdot, \cdot)$ is given by the following:
\begin{equation}
\frac{1}{m ^ \star (m,t)} \triangleq \frac{1}{m} + \frac{2}{t}.
\label{eq:m-star}
\end{equation}
\label{thm:LLED_strict}
\end{theorem}
\begin{IEEEproof}
Recall that 
\begin{align}
\underline{R}^{n,m,t} &= I (X^n; \theta^{(1)}| Y^m).
\end{align}
Further, note the following Markov chain
\begin{equation}
\widehat{\theta^{(1)}}  (X^n) \to \theta^{(1)} \to Z^t \to \theta^{(2)} \to Y^m.
\end{equation}
Assuming that $\min\{m, t\}  = \omega_n(1)$, i.e., both grow unbounded with $n$. Then, we can rely on the asymptotic normality of all of the variables above and noting that $\widehat{\theta^{(1)}}  (X^n)$ is a sufficient statistic for $X^n$, then $\theta^{(1)}$ is Gaussian distributed with mean $\widehat{\theta}(Y^m)$ with variance $ \frac{1}{m^\star} = \frac{2}{t} + \frac{1}{m}$ given $Y^m$. Hence, invoking Lemma~\ref{lem:Gaussian-MI} (appendix) we arrive at the desired result.

For $\min\{m, t\} = O_n(1)$, notice that from Lemma~\ref{lem:Markov-MI} we can deduce that
\begin{align}
I (X^n; \theta^{(1)})  - I (X^n; \theta^{(1)}| Y^m) & = I(X^n; Y^m) \nonumber\\
& \leq I(\theta^{(1)} ; Y^m).
\end{align}
Hence, the result is concluded by noting that $I(\theta^{(1)} ; Y^m) = O_n(1)$.
\end{IEEEproof}

Theorem~\ref{thm:LLED_strict} characterizes the average maximin redundancy in the case of {\LLED} with side information from a correlated source. If the sources are not sufficiently correlated or the side information string is not long enough, then not much performance improvement is expected and the redundancy is close to that of {\LL} strategy. On the other hand, for sufficiently correlated information sources with sufficiently long side information string, one expects that the redundancy would be significantly reduced. In a sense, $m^\star(m,t) $ can be thought of as the effective length of the side information string. When $t \to \infty$, we see that $m^\star(m,t) \approx m$ while for smaller $t$, we see that $m^\star(m,t) <m$. 

\begin{figure}[t]
\centering
\includegraphics[width = 0.46\textwidth]{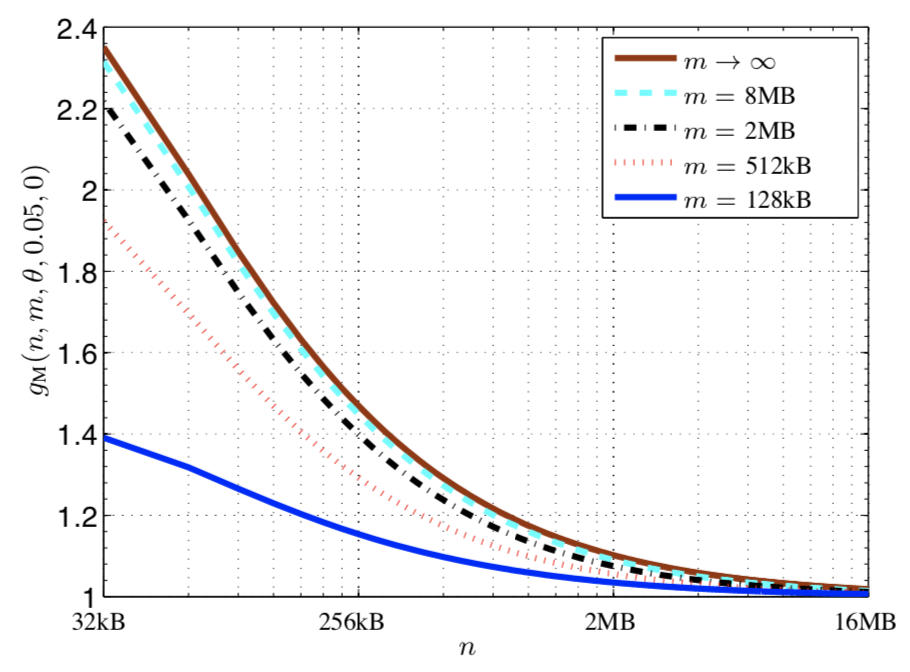}
\caption{Theoretical lower bound on the side information gain~$g_\text{M}(n,\MM,\theta,0.05,0)$ for the first-order Markov source with alphabet size $k=256$ and entropy rate $H^n(\theta)/n = 0.5$.}
\label{fig:M1_256_gain_n}
\end{figure}

\section{Side Information Gain}
\label{sec:gain}
In this section, we define and characterize the side information gain in the different coding strategies described in Section~\ref{sec:setup}. Side information gain is defined as the ratio of the expected codeword length of the traditional universal compression (i.e., {\LL}) to that of the universal compression with side information from a correlated source (i.e., {\LLED}):
\begin{align}
g^{n,m,t} (\theta) &\triangleq \frac{H^n(\theta) + \underline{R}^n}{H^n(\theta) +\underline{R}^{n,m,t}},
\label{eq:def-g}
\end{align}
In other words, $g^{n,m,t}(\theta)$ is  the side information gain on a string of length $n$ drawn from $\mu^n_{\theta}$ and compressed using {\LLED} coding strategy with a side information string of length $m$ drawn from a correlated source with degree of correlation $t$. 

The following is a trivial lower bound on the side information gain.
\begin{lemma}
For all $n,m, t \geq 0$, and $\theta \in \Theta_d$:
\begin{equation}
g^{n,m,t}(\theta) \geq 1.
\end{equation}
\label{thm:g_1_single}
\end{lemma}
\begin{IEEEproof}
This is proved by invoking Lemma~\ref{lem:sandwich}.
\end{IEEEproof}

Next, we present our main result on the side information gain in the next theorem.
\begin{theorem}
If $\min \{ m, t\} = O_n(1)$, then $g^{n,m,t}(\theta) = 1+ O(\frac{1}{n})$. 
If $\min \{m,t\} = \omega_n(1)$:
\begin{equation}
g^{n,m,t} (\theta) = 1 + \frac{\underline{R}^n - \widehat{R}(n, m, t)}{H^n(\theta) + \widehat{R}(n, m, t)}  + O\left(\frac{1}{n}\right) ,
\label{eq:gain-thm}
\end{equation}
where $\widehat{R}(n, m, t)$ is defined in~\eqref{eq:Rhat}.
\label{thm:gain}
\end{theorem}
\begin{IEEEproof}
The theorem is proved by invoking Theorem~\ref{thm:LLED_strict} and light algebraic manipulations.
\end{IEEEproof}

Consider the case where the string length $n$ grows to infinity. Intuitively, we would expect the side information gain to vanish in this case.
\begin{lemma}
For any $m,t \geq 0$ and any $\theta \in \Theta_d$, we have 
\begin{equation}
\lim_{n \to \infty} g^{n,m,t}(\theta) = 1.
\end{equation}
\label{thm:Gain_infinite_n}
\end{lemma}

Let us demonstrate the significance of the side information gain through an example. We let the information source be a first-order Markov source with alphabet size $k=256$. We also assume that the source is such that $H^n(\theta)/n = 0.5$ bit per source character (byte). 
In Fig.~\ref{fig:M1_256_gain_n}, the lower bound on the side information gain is demonstrated as a function of the sequence length $n$ for different values of the memory size $\MM$.
As can be seen, significant improvement in the compression  may be achieved using memorization. For example, the lower bound on $g^{32\text{kB},\MM, \infty}(\theta)$ is equal to $1.39$, $1.92$, $2.22$, and $2.32$, for $\MM$ equal to $128$kB, $512$kB, $2$MB, and $8$MB, respectively. Further, $g^{512\text{kB},\infty,\infty}(\theta) = 2.35$.
Hence, more than a factor of two improvement is expected on top of traditional universal compression when network packets of lengths up to $32$kB are compressed using side information. See~\cite[Section III]{TON14} for practical compression methods that aim at achieving these improvements. 
 As demonstrated in Fig.~\ref{fig:M1_256_gain_n}, the side information gain for memory of size $8$MB is very close to $g^{n,\infty, \infty}(\theta)$, and hence, increasing the memory size beyond $8$MB does not result in substantial increase of the side information gain.
On the other hand, we further observe that as $n\to\infty$, the side information gain becomes negligible regardless of the length of the side information string. For example, at $n=32$MB even when $\MM \to \infty$, we have $g^{32\text{MB},\infty, \infty} \approx 1.01$, which is a subtle improvement. This is not surprising as the redundancy that is removed via the side information is $O(\log n)$, and hence the gain in~\eqref{eq:gain-thm} is $O(\frac{\log n}{n})$ which vanishes as $n$ grows.

Thus far, we have shown that significant performance improvement is obtained from side information on the compression of finite-length strings from low-entropy sources. As also was evident in the previous example, as the size of the memory increases the performance of the universal compression with side information is improved. However, there is a certain memory size beyond which increasing the side information length does not provide further compression improvement. In this section, we will quantify the required size of memory such that the benefits of the memory-assisted compression apply.

Then, the following theorem determines the size of the required memory for achieving $(1-\delta)$ fraction of the gain for unlimited memory.
Let $\widehat{g}^{n, t}(\theta)$ be defined as
\begin{equation}
\widehat{g}^{n, t}(\theta)
\triangleq 1 + \frac{\underline{R}^n}{H^n(\theta)}.
\end{equation}
It is straightforward to see that $\widehat{g}^{n, t}(\theta)$ is the limit of side information gain as the effective side information string length $m^\star(m,t) \to \infty,$ where $m^\star(\cdot, \cdot)$ is defined in~\eqref{eq:m-star}.
\begin{theorem}
Let $m^n_{\delta}(\theta)$ be defined as
\begin{equation}
m_{\delta}^{n} ( \theta) \triangleq \frac{1-\delta}{\delta} \frac{n}{H^n(\theta)} \frac{d}{2} \log e .
\end{equation}
Then, for any $m,t \geq 0$ such that $m^\star (m,t) \geq m_{\delta}^{n} ( \theta)$, we have
\begin{equation}
g^{n,m, t}(\theta) \geq (1-\delta)\widehat{g}^{n, t}(\theta).\nonumber
\end{equation}
\label{thm:m_size}
\end{theorem}
\begin{IEEEproof}
By invoking Theorem~\ref{thm:gain}, we have
\begin{equation}
g^{n,m, t} (\theta) \geq \frac{H^n(\theta)}{H^n(\theta) + \widehat{R}(n,m,t )} \widehat{g}^{n, t}(\theta).
\end{equation}
Hence, we need to show that for $m^\star(m,t) > m_{\delta}^{n, t} ( \theta)$, we have 
\begin{equation}
\frac{H^n(\theta)}{H^n(\theta) + \widehat{R}(n,m,t )} \geq (1-\delta).
\end{equation}

By noting the definition of $m^n_\delta(\theta)$, for any $m^\star(m,t)> m^n_\delta(\theta)$, we have
\begin{equation}
\frac{d}{2}\frac{n}{m^\star(m,t)}\log e \leq \frac{\delta}{1-\delta} H^n(\theta).
\end{equation}
By noting that $\log \left(1+ \frac{n}{m}\right) \leq \frac{n}{m}\log e$, we have
\begin{equation}
\frac{d}{2}\log \left(1+ \frac{n}{m^\star(m,t)}\right) \leq \frac{\delta}{1-\delta} H^n(\theta),
\end{equation}
and hence, the proof is completed by noting the definition of $\widehat{R}(n,m,t)$ in~\eqref{eq:Rhat} and light algebraic manipulations.
\end{IEEEproof}

Theorem~\ref{thm:m_size} determines the size of the memory that is sufficient for the gain to be at least a fraction $(1-\delta)$ of the gain obtained as $m\to \infty$.
Considering our working example of the first-order Markov source in this section with $H^n(\theta)/n = 0.5$, with $\delta = 0.01$, we have
$m_\delta(\theta) \approx 8.9$MB is sufficient for the gain to reach $99\%$ of its maximum confirming our previous observation. This also complements the practical observations reported in~\cite[Section IV.C]{TON14}.

\section{Impact of Prefix Constraint}
\label{sec:one-to-one}
Thus far, all of the results of the paper are on prefix-free codes that satisfy Kraft inequality in~\eqref{eq:Kraft}. However, we remind the reader that our main application is in network packet compression. 
In this case, the code need not be uniquely decodable (satisfy Kraft inequality) as the beginning and the end of each block is already determined by the header of the packet. Thus, the unique decodability condition is too restrictive and can be relaxed. It is only necessary for the mapping (the code) to be injective so as to ensure that {\em one} block of length $n$ can be uniquely decoded. Such codes are known as one-to-one codes. These are also called nonsingular codes in~\cite[Chapter 5.1]{cover-book}.
An interesting fact about one-to-one codes is that while the average codeword length of prefix-free codes can never be smaller than the Shannon entropy, the average codeword length of one-to-one codes can go below the entropy (cf.~\cite{Alon_Orlitsky_one2one,Szpankowski_one2one,Kontoyiannis_one2one,Cover_one2one,Szpankowski2011} and the references therein).

Let $l^n_*(\cdot)$ denote a strictly lossless one-to-one length function. Further, denote $L^n_*$ as the collection of all one-to-one codes (bijective mappings to binary sequences) on sequences of length $n$.
Let $R^n_*(l^n_*, \theta) $ denote the average redundancy of the one-to-one code, which is defined in the usual way as
\begin{equation}
R^n_*(l^n_*, \theta) \triangleq  E\{l^n_*(X^n)\} - H_n(\theta).
\end{equation}
Further, define 
\begin{equation}
\underline{R}^n_* \triangleq \max_{w \in \Omega_d} \min_{l^n_* \in L^n_*} \int_{\theta \in \Theta_d} R^n_*(l^n_*, \theta) w(\theta) d\theta,
\end{equation}
where $\Omega_d$ denotes the set of probability measures on $\Theta_d$.

\begin{theorem}
The following bound holds:
\begin{equation}
\underline{R}^n_*\geq \frac{d-2}{2}\log \frac{n}{2\pi e} -\log 2\pi e^2 + \int_{\theta \in \Theta_d} |\mc{I}(\theta)|^{\frac{1}{2}}d\theta + O\left(\frac{1}{\sqrt{n}}\right).
\end{equation}
\label{thm:one2one_lower}
\end{theorem}
\begin{IEEEproof}
We have
\begin{equation}
H(X^n) = H(X^n|\theta) + I(X^n; \theta)
\end{equation}
Assuming that $\theta$ follows Jeffreys' prior, we can get
\begin{equation}
H(X^n) = \underline{H}^n + \underline{R}^n,
\end{equation}
where $\underline{R}^n$ is the average minimax redundancy for prefix-free codes given in~\eqref{eq:maximin} and $\underline{H}^n$ is given by
\begin{equation}
\underline{H}^n \triangleq \int_{\theta \in \Theta_d} H_n(\theta) w_J(\theta) d\theta.
\end{equation}
We now invoke the main theorem in \cite{Alon_Orlitsky_one2one} to obtain a lower bound on $E \{ l^n_* (X^n)\}$. The proof is completed by observing that $\log \underline{H}^n \leq \log n $ and noting that the average redundancy for the case where $\theta$ follows Jeffreys' prior provides a lower limit on the average maximin redundancy.
\end{IEEEproof}

\begin{figure}[tb]
\centering
\epsfig{width= 0.47\textwidth,file=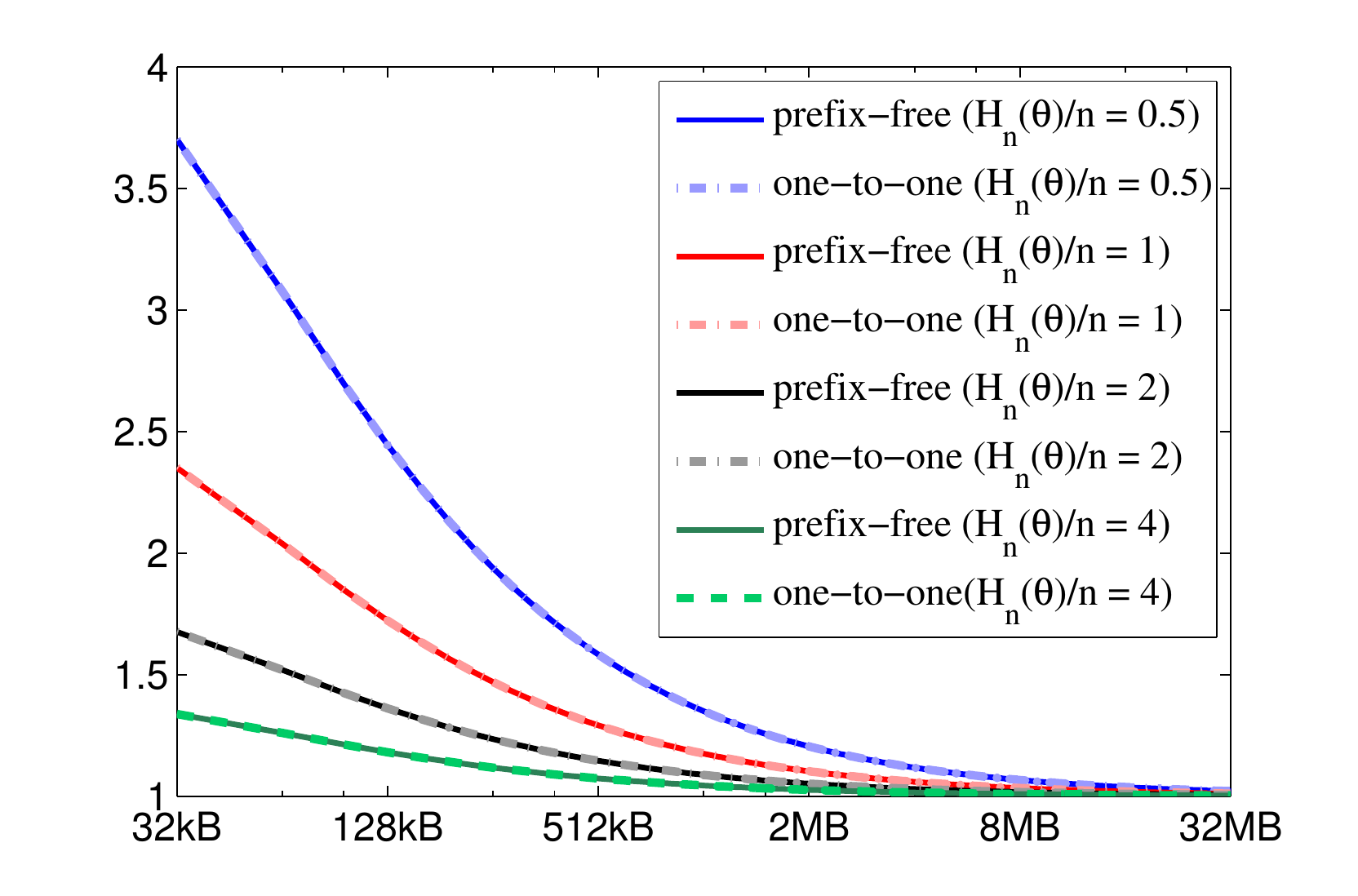}
\caption{Average maximin redundancy as a function of string length for prefix-free and one-to-one universal codes for different values of entropy rate $H_n(\theta)/n$.}
\label{fig:M1_256_one2one}
\end{figure}

Theorem~\ref{thm:one2one_lower} shows that the compression overhead as measured against entropy is $\frac{d-2}{2} \log n + O_n(1)$. However, as discussed earlier, non-universal one-to-one codes achieve an average codeword length that can go below entropy. In particular, for the family of parametric sources studied in this paper, for almost all $\theta \in \Theta_d$, it is shown that the average codeword length is given by $H^n(\theta) - \frac{1}{2} \log n + O_n(1)$~\cite{Szpankowski_one2one,Szpankowski2011,Kontoyiannis_one2one}. Hence, the cost of universality is $\frac{d-1}{2} \log n + O_n(1)$. See~\cite{ITW14, kosut-IT-one2one} for a more complete study of the one-to-one universal compression problem. Additionally, see~\cite{Beirami-IT-Guesswork} for new insights on why the cost of universality scales with one less parameter in one-to-one compression, i.e., $\frac{d-1}{2}$, as compared to $\frac{d}{2}$ for prefix-free codes.

It is desirable to see how much reduction is offered by universal one-to-one compression compared with the prefix-free universal compression. 
We compare the performance of universal one-to-one codes with that of the universal prefix-free codes through the running numerical example from Section~\ref{sec:motivation}.
This example is based on a first-order Markov source with alphabet size $|\mc{X}|=256$, where the number of source parameters is $d = 256 \times 255 = 62580$. 
Note that we have not provided an actual code for the one-to-one universal compression. We compare the converse bound of Theorem~\ref{thm:one2one_lower} with the average maximin redundancy of universal prefix-free  codes.

Fig.~\ref{fig:M1_256_one2one} compares the minimum average number of bits per symbol required to compress the class of the first-order Markov sources normalized to the entropy of the sequence for different values of entropy rates in bits per source symbol (per byte).
 As can be seen, relaxing the prefix constraint at its best does not offer meaningful performance improvement on the compression performance as the curves for the prefix-free codes and one-to-one codes almost coincide. 
 This leads to the conclusion that the universal one-to-one codes are not of much practical interest.

On the other hand, if the source entropy rate is $1$ bit per byte ($H_n(\theta)/n = 1$), the compression rate on sequences of length $32$kB (for both prefix-free and one-to-one codes) is around 2.25 times the entropy-rate, which results in more than 100\% overhead on top of the entropy-rate for both prefix-free and one-to-one universal codes. Hence,  we conclude that average redundancy poses significant overhead in the universal compression of finite-length low-entropy sequences, such as the Internet traffic, which cannot be compensated by dropping the prefix constraint. Hence, the side information gain provided from a correlated information source is essential even if the prefix constraint is dropped.

\begin{figure}[t]
\centering
\includegraphics[height=2.6in, angle=-90]{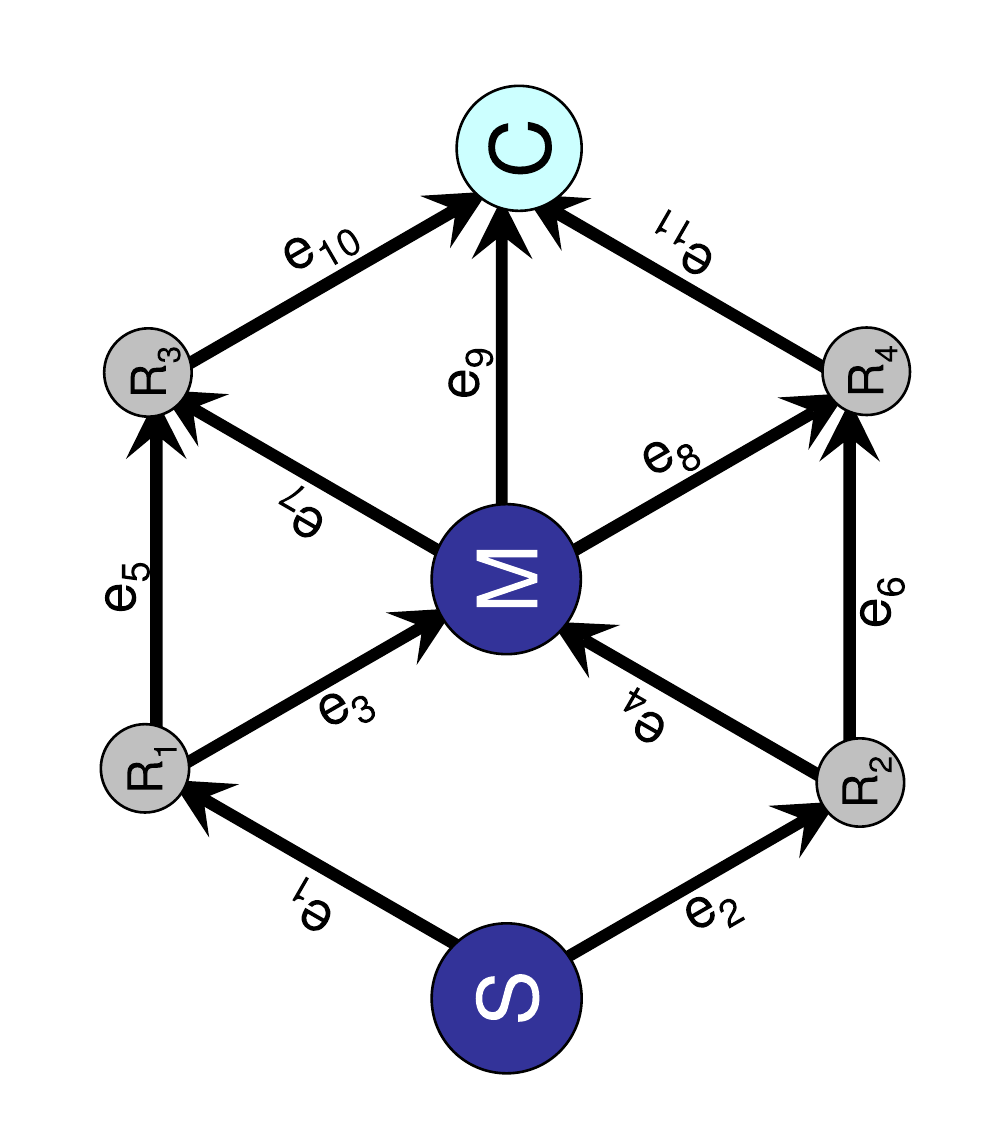}
\vspace{-0.05in}
\caption{The sample network in case study.} 
\label{fig:sample_network_2}
\end{figure}

\section{A Network Case Study}
\label{sec:network}
In this section, we demonstrate how the side information gain could be leveraged in terms of the
compression of network traffic.  Assume that source $S$ is the CNN server and the packet size is $n = 1$kB.
Further, assume that the memory size is $4$MB. In Section~\ref{sec:motivation}, we demonstrated that for this source, the average compression ratio for {\LL} is $\frac{1}{n}E \{ l^n(X^n)\} = 4.42$ bits per byte for this packet size. We further expected that the side information gain for such packet size be at least $g = 5$. Note that the rest of this discussion is concerned as to how the side information gain impacts the overall performance in the network.

We define the network-wide gain of side information measured in
\emph{bit$\times$hop} ($\text{\text{BH}}$) for the
sample network presented in
Fig.~\ref{fig:sample_network_2}, where $M$ denotes the memory element.
Assume that the server $S$ would serve the client $C$ in the network. The intermediate nodes $R_i$ are not capable of memorization.
Recall that the side information gain $g$ is only achievavle on every link in a path where the encoder and the decoder both have access to the side information string.

Let $d(S,C)$ denote the length of the shortest path from $S$ to $C$, which is clearly $d(S,C) = 3$, e.g., using the path $e_1,e_5,e_{10}$.
Let $\text{BH}(S,C)$ denote the minimum  bit-hop cost required to transmit the sequence (of length $n$) from $S$ to $C$ without any compression mechanism, which is $\text{BH}(S,C) = 24$kbits (which is $1$kB$\times8$bits/byte$\times3$). In the case of end-to-end universal compression, i.e., using {\LL}, on the average we need to transmit $\text{BH}_{\LL} = E\{l^n(X^n)\} d(S,C)$
\emph{bit$\times$hop} for the transmission of a packet of length $n$ to the client.

On the other hand, in the case of universal compression with side information, i.e., using {\LLED}, for every information bit on the path from the server to the memory element $M$, we can leverage the
side information, and hence, we only require $\frac{1}{n}E\{ l^{n,m}(X^n, Y^m)\} = \frac{1}{ng} E\{ l^n(X^n)\}$ bit transmissions per each source character that is transmitted to the memory element.
Then, the memory element $M$ will decode the received codeword using {\LLED} decoder and the side information string $y^m$. It will then re-encode the result using {\LL} encoder for the final destination (the client $C$).
In this example, this implies that we require to transmit $\frac{2}{n} E\{ l^{n,m}(X^n, Y^m)\}$ \emph{bit$\times$hop} on
the average from $S$ to $M$ on links $e_1$ and $e_3$ (where $d(S,M) =
2$) for each source character. Then, we transmit the message using $E \{ l^n(X^n)\}$
\emph{bit$\times$hop} per source character from $M$ to $C$ on the link $e_9$. Let
$\text{BH}_{\LLED}$ be the minimum \emph{bit$\times$hop} cost for transmitting the string (of length $n$) using network compression that leverages side information in the $S \to M$ path, i.e., 
\begin{align}
\text{BH}_{\LLED} &= d(S,M) E \{ l^{n,m}(X^n,Y^m)\} \nonumber\\
&+ d(M, C) E \{ l^n(X^n)\}\nonumber\\
& = 2E \{ l^{n,m}(X^n,Y^m)\}+ E \{ l^n(X^n)\}.
\end{align}
Further, let $\mc{G}_{\text{BH}}$ be the
\emph{bit$\times$hop} gain of network compression, defined as $\mc{G}_{\text{BH}} = \frac{\text{BH}_{\LL}}{
\text{BH}_{\LLED}}$. Thus, $\mc{G}_{\text{BH}} = 2.14$ in this example by substituting $g= 5$. In other
words, network compression (using {\LLED} in the $S \to M$ path) achieves more than a factor of $2$ saving in \emph{bit$\times$hop} over the traditional universal compression of the packet (using {\LL} from $S$ to $C$) in the sample network.

In~\cite{TON14}, we fully characterize the scaling of the \emph{bit$\times$hop} gain, $\mc{G}_\text{BH}$, for scale-free networks (random power-law graphs) as a function of side information gain, $g$. We show that $\mc{G}_\text{BH} \approx g$ if the fraction of nodes in the network equipped with memorization capability is larger than a phase-transition cutoff. We refer the interested reader to~\cite{TON14} for more details.

\section{Conclusion}
\label{sec:conclusion}
In this paper, we formulated and studied universal compression with side information from a correlated source.
We showed that redundancy can impose a significant overhead in universal compression of finite-length sequences, such as network packets. We put forth a notion of correlation between information sources where the degree of correlation is controlled by a single hyperparameter. We showed that side information from a correlated source can significantly suppress the redundancy in universal compression. We defined the side information gain and showed that it can be large with reasonable side information size for small strings, such as network packets. 
We showed that this gain is largely preserved even if the code is allowed to be only almost lossless allowing a sufficiently small error that vanishes asymptotically. We also showed that dropping the prefix constraint would not remedy the universal compression problem either. Finally, we showed how these benefits are applicable in network compression in a case study.

\appendix
\begin{assumption}[regularity conditions]
We need some regularity conditions to hold for the parametric model so that our results can be derived.
\begin{enumerate}
\item The parametric model is smooth, i.e., twice differentiable with respect to $\theta$ in the interior of $\Theta_d$ so that the Fisher information matrix can be defined. Further, the limit in~\eqref{eq:fisher} exists.

\item The determinant of fisher information matrix is finite for all $\theta$ in the interior of $\Theta_d$ and the normalization constant in the denominator of~\eqref{eq:Jeffreys} is finite.

\item The parametric model has a minimal $d$-dimensional representation, i.e., $\mc{I}(\theta)$ is full-rank. Hence, $\mc{I}^{-1}(\theta)$ exists.

\item We require that the central limit theorem holds for the maximum likelihood estimator $\widehat{\theta}(x^n)$ of each $\theta$ in the interior of $\Theta_d$ so that $(\widehat{\theta}(X^n) - \theta)\sqrt{n}$ converges to a normal distribution with zero mean and covariance matrix $\mc{I}^{-1}(\theta)$.
\end{enumerate}
\label{assump:reg}
\end{assumption}

\begin{lemma}
For all $t \geq 0$, we have
\begin{align}
E_{\theta^{(1)} \sim q} \left\{  f^t( \theta^{(1)}, \theta^{(2)}) \right\} &= \int_{\theta^{(1)} \in \Theta_d} f^t( \theta^{(1)}, \theta^{(2)}) w(\theta^{(1)}) d\theta^{(1)} \nonumber\\
& = 1.
\label{eq:f-prop1}
\end{align}
\label{lem:f-prop1}
\label{lem:integral}
\end{lemma}
\begin{IEEEproof}
Following the equations
\begin{align}
&\int_{\theta^{(1)} \in \Theta_d} f^t( \theta^{(1)}, \theta^{(2)}) w(\theta^{(1)}) d\theta^{(1)}\nonumber \\
& = \sum_{z^{t} \in \mc{X}^t}{\left(   \frac{\mu^t_{\theta^{(2)}}(z^{t})\int_{\theta^{(1)} \in {\Theta_d}} \mu^t_{\theta^{(1)}}(z^{t}) w(\theta^{(1)}) d\theta^{(1)}}{\int_{\phi \in {\Theta_d}} \mu^t_{\phi}(z^{t}) w(\phi) d\phi}\right)}\nonumber\\
& = \sum_{z^{t} \in \mc{X}^t}\mu^t_{\theta^{(2)}}(z^{t})\label{eq:num-den}\\
& = 1,
\end{align}
where~\eqref{eq:num-den} is obtained since the two integral terms in the numerator and denominator cancel each other out.
\end{IEEEproof}

\begin{lemma}
Let $X \to Y \to Z$ form a Markov chain. Then,
\begin{equation}
I(X;Y) \geq I(X; Y|Z),
\end{equation}
with equality if and only if $I(X;Z) = 0$. The gap in the inequality is also fully characterized by
\begin{equation}
I(X;Y) - I(X;Y|Z) = I(X;Z).
\end{equation}
\label{lem:Markov-MI}
\end{lemma}
\begin{IEEEproof}
This is a well-known result on Markov chains and could be proved by applying the chain rule to $I(X;Y, Z)$ in different orders and noting that $I(X;Z|Y) = 0$ due to the Markov chain.
\end{IEEEproof}

\begin{lemma}
Let $X \to Y \to Z$ form a Markov chain, where $X, Y, Z$ are all Gaussian distributed and supported on $\mathbb{R}^d$. Further, let $X$ follow a non-informative improper uniform distribution on $\mathbb{R}^d$. Let $Y$ be a noisy observation of $X$ with variance $\sigma^2,$ i.e., $Y = X + N_1$ where $N_1\sim \mc{N}(\mb{0}_d, \sigma^2 I_d)$ is independent of $X$, and $\mb{0}_d$ and $I_d$ denote the $d$-dimensional all-zero vector and identity matrix, respectively. In the same way, let $Z$ be a noisy observation of $Y$ with variance $\tau^2$. Then,
\begin{equation}
I(X;Y|Z) = \frac{d}{2} \log \left( 1 + \frac{\tau^2}{\sigma^2} \right) .
\end{equation}
\label{lem:Gaussian-MI}
\end{lemma}
\begin{IEEEproof}
The proof is completed by following the following equations:
\begin{align}
I(X;Y|Z) &= h(X|Z) - h(X|Y, Z) \\
& = h(X|Z) - h(X|Y)\\
& = \frac{d}{2} \log \left(2\pi e (\sigma^2+ \tau^2)\right) - \frac{d}{2} \log (2\pi e \sigma^2)\\
& = \frac{d}{2} \log \left( 1 + \frac{\tau^2}{\sigma^2} \right).
\end{align}
where $h$ denotes differential entropy.
\end{IEEEproof}

\begin{lemma}
The following inequality holds:
\begin{equation}
H(X^n|\mb{1}_e(X^n), \widehat{X}^n) \leq \epsilon H(X^n).
\end{equation}
\label{lem:ineq_entrop}
\end{lemma}
\begin{IEEEproof}
\begin{eqnarray}
H(X^n|\mb{1}_e(X^n), \widehat{X}^n) \hspace{-0.09in}&= \hspace{-0.09in}& (1-\epsilon)H(X^n|\mb{1}_e(X^n,) = 0 , \widehat{X}^n)\nonumber\\
 \hspace{-0.09in}&+ \hspace{-0.09in}& \epsilon H(X^n|\mb{1}_e(X^n) = 1, \widehat{X}^n)\label{eq:22}\\
\hspace{-0.09in}&\leq \hspace{-0.09in}& \epsilon H(X^n)\label{eq:23}.
\end{eqnarray}
The first term in (\ref{eq:22}) is zero since if $\mb{1}_e(X^n) = 0$, we have $X^n = \widehat{X}^n$ and hence $$H(X^n|\mb{1}_e(X^n,) = 0 , \widehat{X}^n) = 0.$$ The inequality in~(\ref{eq:23}) then follows from the fact that conditioning does not increase entropy, completing the proof.
\end{IEEEproof}

\begin{lemma}
The average minimax redundancy for the almost lossless {\LL} coding strategy is lower bounded by
\begin{equation}
\underline{R}^n_\epsilon \geq (1-\epsilon)\underline{R}^n - h_b(\epsilon) - \epsilon H^n(\theta),\nonumber
\end{equation}
where $h_b(\epsilon)$ is the binary entropy function defined as:
\begin{equation}
h_b(\epsilon) \triangleq \epsilon \log\left(\frac{1}{\epsilon}\right) + (1-\epsilon) \log\left(\frac{1}{1-\epsilon}\right).
\label{eq:binary_entropy}
\end{equation}
\label{thm:NM_almost}
\end{lemma}
\begin{IEEEproof}
Consider $H(X^n, \widehat{X}^n, \mb{1}_e(X^n))$. Note that both $\widehat{X}^n$ and $\mb{1}_e(X^n)$ are deterministic functions of $X^n$ and hence
\begin{equation}
H(X^n, \widehat{X}^n, \mb{1}_e(X^n)) = H(X^n).
\end{equation}
On the other hand, we can also use the chain rule in a different order to obtain
\begin{eqnarray}
H(X^n, \widehat{X}^n, \mb{1}_e(X^n)) = H(\widehat{X}^n) + H(\mb{1}_e(X^n) | \widehat{X}^n)\nonumber\\
 + H(X^n|\mb{1}(X^n) , \widehat{X}^n).
\end{eqnarray}
Hence,
\begin{eqnarray}
H(\widehat{X}^n)&\hspace{-0.09in} = \hspace{-0.09in}& H(X^n) - H(\mb{1}_e(X^n) | \widehat{X}^n) - H(X^n|\mb{1}(X^n) , \widehat{X}^n)\nonumber\\
&\hspace{-0.09in}\geq \hspace{-0.09in}& H(X^n) - h_b(\epsilon) - H(X^n|\mb{1}(X^n), \widehat{X}^n)\label{eq:ineq1}\\
&\hspace{-0.09in}\geq \hspace{-0.09in}& H(X^n) - h_b(\epsilon) - \epsilon H(X^n)\label{eq:lem_ineq},
\end{eqnarray}
where the inequality in~(\ref{eq:ineq1}) is due to the fact that $H(\mb{1}_e(X^n) | \widehat{X}^n) \leq H(\mb{1}_e(X^n)) = h_b(\epsilon)$ and the inequality in~(\ref{eq:lem_ineq}) is due to Lemma~\ref{lem:ineq_entrop}.
The proof of the theorem is completed by noting that $H(X^n) = H^n(\theta) + \underline{R}^n$.
\end{IEEEproof}

\ifCLASSOPTIONcaptionsoff
  \newpage
\fi

\bibliographystyle{IEEEtran}
\bibliography{phd_thesis_bib}

\end{document}